\documentclass[10pt,conference]{IEEEtran}
\usepackage{cite}
\usepackage{balance}
\usepackage{amsmath,amssymb}
\usepackage{graphicx}
\usepackage{booktabs,array,tabularx,multirow}
\usepackage[table]{xcolor}
\usepackage{listings}
\usepackage{tikz}
\usetikzlibrary{arrows.meta,positioning,calc}
\usepackage{pgfplots}
\pgfplotsset{compat=1.17}
\usepackage[hidelinks]{hyperref}
\usepackage{url}

\newcommand{\designmd}{\texttt{DESIGN.md}}
\newcommand{\dataset}{DesignSpecs}
\newcommand{\code}[1]{\texttt{#1}}
\newcommand{\papertitle}{DesignSpecs: A DESIGN.md Dataset of Visual Design Specifications for AI Agents}
\newcommand{\paperkeywords}{DESIGN.md, Google Stitch, Design Systems, User Interface, AI Agents, Mining Software Repositories, AI-Inspired UI Design}
\hypersetup{pdftitle={\papertitle},pdfauthor={Jialiang Wei and Walid Maalej},pdfkeywords={\paperkeywords}}

\definecolor{accent}{HTML}{2F5D8A}
\definecolor{boxfill}{HTML}{EEF3F8}
\lstdefinelanguage{designmd}{morecomment=[l]{\#\#},morestring=[b]",sensitive=true}
\begin{document}
\title{\papertitle}

\author{\IEEEauthorblockN{Jialiang Wei}
\IEEEauthorblockA{\textit{Hasso Plattner Institute} \\
\textit{University of Potsdam}\\
Potsdam, Germany \\
jialiang.wei@hpi.de}
\and
\IEEEauthorblockN{Walid Maalej}
\IEEEauthorblockA{\textit{Hasso Plattner Institute} \\
\textit{University of Potsdam}\\
Potsdam, Germany \\
walid.maalej@hpi.de}}

\maketitle

\begin{abstract}
While AI agents increasingly generate and modify user interfaces, they rarely rely on a visual design specification for a certain project.
The \designmd{} format, which originated from Google's AI design tool Stitch and which was open-sourced in April 2026, addresses this gap by describing a visual identity through design tokens (such as named colors, text styles) and Markdown prose.
To our knowledge, no large-scale research dataset of such design specification files exists.
We present \dataset{}, a dataset of 36,906 \designmd{} files from 15,794 public, non-fork GitHub repositories, representing 17,240 unique file contents.
We collected 613,207 files named \designmd{} in October 2026. 
Because the same file name also denotes architectural plans and other design documents, we filtered those with YAML front matter, a design-system name, at least one recognized design token, and zero errors under the official linter.
Our JSON Lines dataset provides file-level and deduplicated views of the original Markdown, links files to repositories, paths, and commits, and is released with related scripts.
We outline research opportunities on visual design knowledge and design-aware AI agents.
\end{abstract}


\section{Introduction}
A visual design shapes how users understand software: colors convey meaning, typography establishes hierarchy, and consistent spacing and components create a coherent interface.
Design systems capture these choices through guidelines and \emph{design tokens}: named, typed values, such as a \code{primary} color or a \code{body-md} text style, that can be shared between design tools and code~\cite{dtcg2025format}.
In many repositories, however, these decisions remain implicit, scattered across CSS variables, framework themes, and screenshots.

AI agents generate and modify front-end code from requests, screenshots, or mock-ups~\cite{Wei:AIInspiredUserInterface:2025,Si:Design2CodeBenchmarkingMultimodal:2025}, which may leave parts of the intended visual identity unspecified.
Context files such as \code{AGENTS.md}, \code{CLAUDE.md}, and \code{SKILL.md} provide project instructions and workflows~\cite{chatlatanagulchai2025manifests,chatlatanagulchai2026agentreadmes,Destefanis:GitSkillsDatasetAgent:2026}.
Few of these files describe how the UI should look.

Google's \designmd{} format makes this visual intent explicit.
It originated in the AI design tool Stitch~\cite{nallatamby2025stitch}, and its draft specification was open-sourced on April 21, 2026~\cite{xu2026stitch,google2026designmd}.
A file combines machine-readable YAML design tokens with human-readable Markdown prose (Fig.~\ref{fig:example}).
Tokens give agents exact values, such as colors, typography, spacing, rounded corners, and component styles; prose explains how and why to apply them.
The official command-line tool checks files, compares versions, and exports tokens to Tailwind and the W3C token format~\cite{google2026designmd}.

\begin{figure}[t]
\begin{lstlisting}[language=designmd]
---
name: Paws & Paths
colors:                             # 47 color roles, e.g.
  primary: "#855300"
  on-primary: "#ffffff"
  primary-container: "#f59e0b"
typography:                         # 8 text styles, e.g.
  headline-lg: {fontFamily: Plus Jakarta Sans, ...}
rounded: {sm: 0.25rem, DEFAULT: 0.5rem, lg: 1rem}
spacing: {xs: 4px, sm: 12px, md: 24px, gutter: 16px}
components:                         # 10 components, e.g.
  button-primary:
    backgroundColor: "{colors.primary}"
    textColor: "{colors.on-primary}"
    rounded: "{rounded.lg}"
---
## Brand & Style
The design system is built to evoke the joyful energy of a
walk in the park balanced with the reliability of a ...
## Colors
The palette centers on "Golden Retriever" orange to ...
\end{lstlisting}
\caption{Excerpt of \designmd{} from the official examples~\cite{google2026designmd}: YAML tokens in the front matter, rationale in Markdown prose.}
\label{fig:example}
\vspace{-4mm}
\end{figure}

\designmd{} files allow the mining software repositories (MSR) community to study visual design knowledge as structured text.
Building such a dataset requires more than a filename search.
Kiro and OpenSpec use \designmd{} for technical architecture and implementation plans~\cite{kiro2026featurespecs,openspec2026}; other projects also use it for general design notes.

\begin{figure*}[t]
\centering
\begin{tikzpicture}[
  stage/.style={draw=accent,fill=boxfill,rounded corners=2pt,align=center,font=\scriptsize,minimum height=14mm,text width=34mm,inner sep=2pt},
  arr/.style={-{Stealth[length=1.6mm]},thick,accent},
  num/.style={font=\scriptsize\bfseries,text=accent}]
\node[stage] (s1) {\textbf{1 Filename search}\\discover repositories and retrieve \designmd{} files at pinned commits};
\node[stage,right=6mm of s1] (s2) {\textbf{2 Header screening}\\retain candidates with \\ a potential YAML front-matter delimiter};
\node[stage,right=6mm of s2] (s3) {\textbf{3 Structural filtering}\\require a header, name,\\recognized tokens, and\\zero official lint errors};
\node[stage,right=6mm of s3] (s4) {\textbf{4 Packaging}\\export three JSONL files;\\verify hashes and links;\\re-run the official linter};
\foreach \a/\b in {s1/s2,s2/s3,s3/s4} \draw[arr] (\a) -- (\b);
\node[num,below=1mm of s1,align=center] {613,207 files saved\\196,585 repositories discovered};
\node[num,below=1mm of s2,align=center] {126,345 candidate files\\31,378 repositories};
\node[num,below=1mm of s3,align=center] {36,909 eligible files\\15,794 repositories};
\node[num,below=1mm of s4,align=center] {36,906 stored files\\17,240 unique contents\\15,794 repositories};
\end{tikzpicture}
\vspace{-1mm}
\caption{The four-stage pipeline for constructing \dataset{}, with file and repository counts at each stage.}
\label{fig:pipeline}
\vspace{-3mm}
\end{figure*}
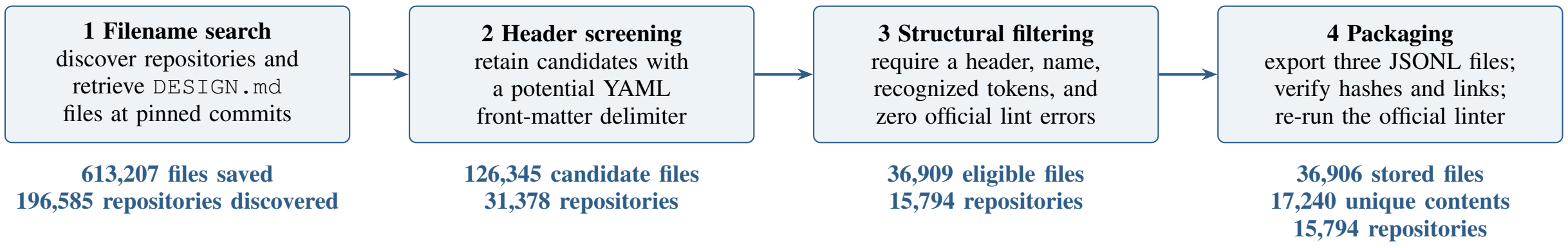

We present \dataset{}, a research dataset dedicated to token-bearing \designmd{} specifications, i.e., files that contain at least one design token recognized by the official linter.
We distinguish \emph{repository files}, each identified by its repository, path, and pinned commit, from \emph{unique file contents} (short: unique contents) after byte-level deduplication.
Our contributions are:
\begin{itemize}
\item a JSON Lines dataset of \textbf{36,906} repository files (\textbf{17,240} unique contents) from \textbf{15,794} public, non-fork GitHub repositories, with original contents, per-file provenance, and repository metadata;
\item scripts to collect and filter the data; and
\item characterization of the dataset and the research it enables.
\end{itemize}

\section{Related Work}
\textbf{Agent context files.}
Chatlatanagulchai et al.\ analyzed the structure and content of \code{CLAUDE.md} manifests~\cite{chatlatanagulchai2025manifests} and later compared context files across agentic tools, including their maintenance over time~\cite{chatlatanagulchai2026agentreadmes}.
GitSkills collects 3.8 million \code{SKILL.md} files with 1.9 million unique contents~\cite{Destefanis:GitSkillsDatasetAgent:2026}; we likewise distinguish repository files from unique contents.
Markdown Registry tracks agent Markdown files more broadly~\cite{registry2026state} and audited 1,162 \designmd{} files with the official linter, of which only 254 contained YAML tokens~\cite{registry2026designaudit}.
\dataset{} extends this work with a larger token-bearing collection, original file contents, traceable provenance, and the scripts used to build it.

\textbf{Spec-driven development.}
SpecMine collects 470,795 \code{spec.md}/\code{specs.md} files plus Kiro and OpenSpec artifacts, including 32,123 Kiro and 79,459 OpenSpec \designmd{} files~\cite{Agarwal:SpecDrivenDevelopmentPrescribed:2026}.
In these tools, \designmd{} documents technical architecture and implementation decisions~\cite{kiro2026featurespecs,openspec2026}.
Their role differs from that of visual specifications, motivating our content-based filtering.

\textbf{UI design datasets.}
Data-driven UI research relies on datasets such as Rico~\cite{Deka:RicoMobileApp:2017} and MUD~\cite{Feng:MUDLargeScaleNoiseFiltered:2024}, which provide screenshots and view hierarchies of Android apps.
GUing mined app introduction images from Google Play and released 303k app screenshots, including 135k image--caption pairs in SCapRepo, to support GUI retrieval~\cite{Wei:GUingMobileGUI:2025}.
Prior work has explored how such resources can support AI-inspired UI design~\cite{Wei:BoostingGUIPrototyping:2023,Wei:AIInspiredUserInterface:2025,Feng:GalleryDCAutocreated:2022,Kolthoff:DatadrivenPrototypingNaturallanguagebased:2023}.
Screenshot-to-code corpora and benchmarks cover web front-ends~\cite{Si:Design2CodeBenchmarkingMultimodal:2025}.
These resources provide examples of UI appearance and implementation.
\dataset{} complements them with statements stored in the repository that articulate the design intent behind a UI.

\section{Dataset Construction}
Fig.~\ref{fig:pipeline} illustrates the pipeline.
We first collect files by name, then filter those that satisfy the rules detailed in the following section.

\subsection{Stage 1: Filename Search}
We searched GitHub for files whose basename equals \designmd{}, ignoring case, using the REST code search API~\cite{github2026search}.
Each query returns at most 1,000 results, so the collector recursively split queries into smaller file-size ranges; searches for repositories whose READMEs mention the specification supplemented these queries.
The searches identified 617,180 file paths across 196,585 repositories.
We collected files only from public, non-fork repositories.
For each repository, we recorded the head commit of its default branch (the \emph{pinned commit}) and retrieved the discovered files at that commit during October 2026.
Of these paths, 3,752 could not be resolved at the pinned commits, and 221 could not be retrieved because the repository API returned HTTP 404.
We ultimately downloaded 613,207 files.

\subsection{Stage 2: Header Screening}
A filename match does not establish that a file contains a visual specification.
We used a lightweight prefix check to identify files that might contain YAML front matter, retaining 126,345 candidates from 31,378 repositories.

\subsection{Stage 3: Structural Filtering}
We evaluated each candidate with the public \code{lint()} function of \code{@google/design.md} 0.4.0, using its default rules.
A file was eligible only if it met all four conditions:
\begin{enumerate}
\item a complete YAML front-matter block bounded by exact \code{---} lines;
\item a nonempty design-system \code{name};
\item at least one recognized design token; and
\item successful linting with zero errors.
\end{enumerate}
Recognized tokens span five groups: colors, typography, spacing, rounded corners, and components.
The header is required, but tokens may also appear in additional fenced YAML blocks.
Linter warnings do not disqualify a file.

We excluded 2,690 candidates without an exact header and 12,972 with linter errors.
A further 73,774 lacked a name, recognized tokens, or both.
This left 36,909 eligible files from 15,794 repositories, which proceeded to packaging.

\subsection{Stage 4: Packaging}
Three eligible paths were lost in an intermediate export to a case-insensitive file system, because each differed from another path in the same repository only in letter case.
Two of them are byte-identical to their retained counterparts; one differs.
These exclusions leave 36,906 stored repository files (Fig.~\ref{fig:pipeline}).

We verified SHA-256 hashes, repository/commit bindings, and source URLs.
Re-running the linter on all 17,240 unique contents reproduced the collection-time results in every case.
The released file records include these diagnostics, allowing users to select stricter subsets, such as warning-free files.

\section{Dataset Schema and Characteristics}
\subsection{Schema and Access}
The three JSON Lines files total approximately 1,115\,MB (Table~\ref{tab:schema}).
\code{files.jsonl} contains each file's full Markdown, source, and linter diagnostics; \code{contents.jsonl} stores each distinct content once.
The \code{sha256} field links these views, and \code{repo\_id} links files to repositories.
The tuple (\code{repo\_id}, \code{commit\_sha}, \code{path}) identifies a file.
Encoding \code{content} as UTF-8 reconstructs the original bytes.
Repository \code{topics} is a string array; \code{languages} maps languages to code byte counts.
\code{schema.json} defines the fields, \code{statistics.json} provides aggregate results, and the README documents the repository structure.
The package includes \code{collect.py} and \code{filter.py} for collection and structural filtering.

\begin{table}[t]
\caption{Overview of the released JSONL files.}
\label{tab:schema}
\vspace{-1mm}
\centering\footnotesize
\begin{tabularx}{\columnwidth}{@{}p{.235\columnwidth}@{\hspace{3pt}}r@{\hspace{5pt}}>{\raggedright\arraybackslash}X@{}}
\toprule
File & Records & Information\\\midrule
\code{contents} & 17,240 & Unique Markdown contents, identified by SHA-256.\\\addlinespace[2pt]
\code{files} & 36,906 & Original Markdown, file provenance, and official linter diagnostics.\\\addlinespace[2pt]
\code{repositories} & 15,794 & Repository metadata, including descriptions, topics, stars, licenses, languages, and dates.\\\bottomrule
\end{tabularx}
\vspace{-3mm}
\end{table}

\subsection{Corpus Overview}
\textbf{Corpus size and duplication.}
The corpus contains 36,906 repository files from 15,794 repositories, representing 17,240 unique file contents.
Copies beyond the first instance of each unique content account for 19,666 files (53.3\%).
Of the unique contents, 842 appear in more than one repository, together accounting for 17,430 files.
The most widely shared content is named \emph{Apple-design-analysis} in its YAML header; byte-identical copies appear in 208 repositories.
Since file-level statistics weight repeated contents, we report unique-content statistics unless stated otherwise.

\textbf{Files per repository.}
Most repositories contribute only one file (13,692; 86.7\%).
The largest repository contributes 1,045 files (2.8\% of all files), and the top 100 repositories together contribute 25.1\%.

\textbf{File names and locations.}
The uppercase spelling \designmd{} accounts for 93.3\% of files.
Nearly half of the repositories (47.3\%) contain a matching file at their root.
A directory named \code{skills} appears in the paths of 20.6\% of files.
Common top-level directories include \code{.agents/} (2,361 files) and \code{.claude/} (1,811 files).

\begin{figure}[t]
\centering
\begin{tikzpicture}
\begin{axis}[
  width=\columnwidth, height=3.3cm, ybar, bar width=7pt,
  ymin=0, ymax=2500, enlarge x limits=0.05,
  symbolic x coords={25-10,25-11,25-12,26-01,26-02,26-03,26-04,26-05,26-06,26-07,26-08,26-09,26-10},
  xtick=data, xticklabels={O,N,D,J,F,M,A,M,J,J,A,S,O*},
  tick label style={font=\scriptsize}, label style={font=\scriptsize},
  ylabel={repositories}, ylabel shift=-3pt,
  ytick={0,1000,2000}, axis x line*=bottom, axis y line*=left,
  xlabel={creation month (October 2025 -- October 2026)}, xlabel shift=-2pt]
\addplot[fill=accent!60,draw=accent] coordinates {(25-10,82)(25-11,113)(25-12,149)(26-01,196)(26-02,293)(26-03,458)(26-04,1013)(26-05,2270)(26-06,2172)(26-07,2184)(26-08,2090)(26-09,2343)(26-10,1352)};
\draw[densely dashed,gray,xshift=1pt] (axis cs:26-04,0) -- (axis cs:26-04,2400);
\node[font=\tiny,align=right,anchor=east] at ([xshift=0pt]axis cs:26-04,1900) {DESIGN.md open-\\sourced (Apr 21)};
\end{axis}
\end{tikzpicture}
\vspace{-2mm}
\caption{Creation months of the 14,715 repositories created since October 2025 (1,079 earlier ones omitted).
*Partial month: collection ended on October 22, 2026.}
\label{fig:created}
\vspace{-3mm}
\end{figure}
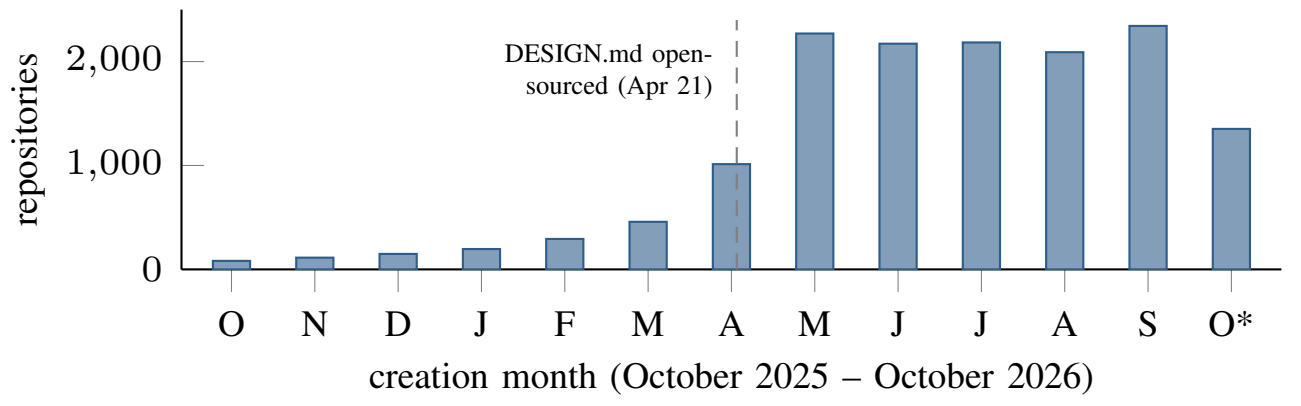

\textbf{Repository characteristics.}
Most repositories were created in 2026 (91.0\%; Fig.~\ref{fig:created}).
The largest monthly cohorts were created after the specification's release, but these dates do not indicate when a \designmd{} file was added.
At collection time, 77.5\% of repositories had no stars, while 160 had at least 100, including Langflow, Ant Design, and marimo.
The most common primary languages are TypeScript (5,986 repositories), HTML (2,952), JavaScript (2,231), and Python (1,803).

\subsection{What the Specifications Contain}

\begin{table}[t]
\caption{Token groups in 17,240 unique contents.
Coverage is the percentage containing each group.
Name frequencies use all unique contents as the denominator.}
\label{tab:groups}
\vspace{-1mm}
\centering\footnotesize
\setlength{\tabcolsep}{3pt}
\begin{tabularx}{\columnwidth}{@{}lr>{\raggedright\arraybackslash}X@{}}
\toprule
Group & Coverage & Common names\\\midrule
Colors & 99.5\% & \code{primary} 78\%, \code{surface} 68\%\\
Typography & 97.9\% & \code{body-md} 56\%, \code{body-lg} 46\%\\
Spacing & 90.2\% & \code{md} 45\%, \code{sm} 45\%, \code{lg} 44\%\\
Rounded & 91.2\% & \code{sm} 77\%, \code{md} 76\%, \code{lg} 72\%\\
Components & 45.3\% & \code{button-primary} 33\%, \mbox{\code{card} 13\%}\\\bottomrule
\end{tabularx}
\vspace{-3mm}
\end{table}

\textbf{Token coverage.}
Table~\ref{tab:groups} summarizes the five token groups recognized by the linter's parser.
Token names and font families are extracted from the YAML headers, so tokens that appear only in fenced blocks are not included in these name counts.
Design information expressed only in prose, such as elevation, motion, or responsive behavior, is outside these counts.
Colors and typography appear in nearly all unique contents, while fewer than half contain component tokens.
Most contents define four (50.5\%) or all five (39.4\%) groups.
Among those with four groups, 90.2\% omit components, retaining colors, typography, spacing, and rounded corners.
The median token count is 65.

\textbf{Token names and values.}
Color names commonly describe roles, such as \code{primary}, \code{surface}, and \code{on-primary}.
The text of 46.5\% of contents includes both \code{on-primary-container} and \code{surface-container}, names used by Material Design 3 and the official examples~\cite{material2026roles}.
Spacing and rounded-corner tokens often use size labels such as \code{sm}, \code{md}, and \code{lg}.
Shared names need not imply identical values: for example, \code{spacing.md} is \code{16px} in 4,069 contents and \code{12px} in 1,599.

Typography names commonly indicate text roles, such as \code{display}, \code{headline}, \code{body}, and \code{label}.
Counting the first family in each token's \code{fontFamily} declaration, Inter appears in 35.2\% of contents, followed by JetBrains Mono (18.7\%), Plus Jakarta Sans (8.8\%), and Geist (6.4\%); percentages overlap.

Component names are dominated by buttons: among contents with component tokens, 72.4\% define \code{button-primary}, equivalent to 32.8\% of all contents.
Related names such as \code{button-primary-hover} and \code{button-primary-disabled} describe interaction states as separate entries.

\textbf{Prose organization.}
The official specification defines eight canonical Markdown sections~\cite{google2026designmd}.
A unique content has a median of seven sections.
We matched headings to the specification's section names and aliases, including ``Brand \& Style'' as Overview.
Typography appears in 90.5\% of contents, Overview in 90.4\%, Colors in 89.2\%, and Components in 88.8\%.
Elevation \& Depth (85.7\%), Layout (82.2\%), and Shapes (76.6\%) are also common.
``Do's and Don'ts'' is the least common of these sections (43.7\%); 31.1\% of contents contain all eight specified headings.
Additional headings include \emph{Responsive Behavior} (7.2\%), \emph{Agent Prompt Guide} (4.6\%), and \emph{Motion} (3.2\%).

Headings do not imply structured tokens: 8,181 contents (47.5\%) have a Components heading but no recognized component tokens.

\textbf{Linter diagnostics.}
All selected files pass the linter without errors.
However, 49.1\% of unique contents and 68.7\% of repository files contain warnings.
For the results below, each unique content is counted once per warning type, regardless of how many times that warning occurs.
Warning categories can overlap.

Among the 7,817 unique contents that define component tokens, 88.8\% include color tokens that are not referenced by any declared component, 43.2\% trigger a contrast warning for at least one component text/background pair below 4.5:1~\cite{wcag22}, and 30.6\% use unrecognized component properties.
These percentages use the 7,817 contents as their denominator.
The color-reference check concerns only references within the specification: an unreferenced color may still be used in the application.
Likewise, the contrast check examines resolvable token values rather than rendered interfaces, so it does not establish interface accessibility.

Two other common warnings are reported across all 17,240 unique contents:
21.5\% lack a \code{primary} color, and 12.0\% use typography properties that the linter does not recognize.

\section{Research Opportunities}
We outline five possible research questions.

\textbf{RQ1: What design knowledge do developers make explicit for agents?}
Token counts and section headings show which decisions are encoded as structured values and which remain in prose; for example, 47.5\% of contents describe components only in prose.
Comparing the vocabulary with Material Design 3 and the W3C token format~\cite{material2026roles,dtcg2025format} can reveal emerging conventions and gaps, such as rarely specified motion or responsive behavior.

\textbf{RQ2: Where do specifications come from, and how do they spread?}
Byte-identical copies account for 53.3\% of files, brand-inspired catalog entries recur in up to 208 repositories, and 20.6\% of files are packaged inside agent skill directories.
Clone and near-duplicate analysis~\cite{lopes2017dejavu} can reveal shared content and variations across specifications; determining how they were produced and the direction of reuse requires additional evidence, such as file histories and generation records.

\textbf{RQ3: Is agent-era visual design converging?}
Inter appears in 35.2\% of contents and 46.5\% reuse Material Design 3 color roles.
Comparing palettes, type scales, and font choices across repositories and over time can test whether shared templates and generators homogenize UI design.

\textbf{RQ4: How do design specifications fit into the agent context ecosystem?}
Many \designmd{} files ship with agent configurations: \code{.agents/} and \code{.claude/} are among their most common top-level directories, and many files live inside agent skills.
Joining the repositories with their \code{AGENTS.md}, \code{CLAUDE.md}, and \code{SKILL.md} files~\cite{chatlatanagulchai2026agentreadmes,Destefanis:GitSkillsDatasetAgent:2026} can show how visual guidance is divided, duplicated, or contradicted across context files, and which agent setups adopt it.

\textbf{RQ5: Do specifications help AI agents build conforming UI?}
Each file's pinned commit allows retrieving the corresponding implementation to measure design--implementation drift, e.g., between tokens and CSS variables or Tailwind themes.
The files can also seed benchmarks that compare UI generation~\cite{Si:Design2CodeBenchmarkingMultimodal:2025} with and without a \designmd{} file, or derive specifications from screenshots in GUI datasets~\cite{Feng:MUDLargeScaleNoiseFiltered:2024,Wei:GUingMobileGUI:2025}, and explore more personalized UI generation~\cite{Wei:LikeThisEmpoweringApp:2026}.

\section{Limitations}
\textbf{Coverage.}
GitHub code search only indexes default branches and files below a size limit, and it returns incomplete results for some queries~\cite{github2026search}; the collected files therefore represent an incomplete view of public GitHub content.
Like other GitHub-based datasets, \dataset{} contains personal and experimental projects~\cite{kalliamvakou2014promises}, and forks are excluded.

\textbf{Construct validity.}
The filtering predicate is structural and tied to \code{@google/design.md} linter version 0.4.0.
It excludes prose-only files and files without YAML front matter, even when their fenced blocks contain usable tokens.
Satisfying the filtering rule does not certify design quality or identify whether a file was written by a person or generated by Stitch.
Some files are catalog entries or templates rather than specifications of a deployed product.

\textbf{Ethics and privacy.}
We collected no commit-author or contact records, but the unchanged public file contents may contain personal information.

\textbf{Temporal scope.}
The snapshot covers one commit per repository, collected during October 2026, without file histories or implementation snapshots.

\section{Data Availability}
\dataset{} and the related scripts are available for review at \url{https://huggingface.co/datasets/Jl-wei/DesignSpecs}.
Upon acceptance, we will archive a versioned release with a DOI on Zenodo.

\section{Conclusion}
We presented \dataset{}, 36,906 \designmd{} files (17,240 unique contents) from 15,794 GitHub repositories.
We hope it will support research on visual design knowledge, design-aware AI agents, and AI-inspired UI design.

\balance
\bibliographystyle{IEEEtran}
\bibliography{ref}
\end{document}